\documentclass{ifacconf}

\usepackage{graphicx}
\usepackage{amssymb,amsmath,bm}
\usepackage[sort]{natbib}

\newtheorem{problem}{Problem}

\usepackage{color}
\definecolor{dgreen}{rgb}{0,.6,0}
\definecolor{purple}{rgb}{.6,0,.6}
\newcommand\modified[1]{\textcolor{dgreen}{#1}} 

\usepackage[normalem]{ulem}

\begin{document}

\begin{frontmatter}
\title{Some Hints for the Design of Digital Chaos-Based Cryptosystems: Lessons Learned from Cryptanalysis}

\thanks[footnoteinfo]{The work described in this paper was supported by
\textit{Minis\-terio de Educaci\'on y Ciencia of Spain}, research
grant SEG2004-02418, \textit{CDTI, Minis\-terio de Industria,
Turismo y Comercio of Spain} in collaboration with Telef\'onica I+D,
Project SEGUR@ with reference CENIT-2007 2004, \textit{CDTI,
Minis\-terio de Industria, Turismo y Comercio of Spain} in
collaboration with SAP, project HESPERIA (CENIT 2006-2009), and
\textit{Ministerio de Ciencia e Innovaci\'on of Spain} in
collaboration, project CUCO (MTM2008-02194). Shujun Li was supported
by a fellowship from the Zukunftskolleg of the Universit\"at
Konstanz, Germany.}

\author[First]{David Arroyo},
\author[First]{Gonzalo Alvarez} and
\author[Third]{Shujun Li}

\address[First]{Instituto de F\'{\i}sica Aplicada, Consejo Superior de
Investigaciones Cient\'{\i}ficas, Serrano 144, 28006 Madrid, Spain
(e-mail: david.arroyo@iec.csic.es).}
\address[Third]{Fachbereich Informatik und Informationswissenschaft, Universit\"at
Konstanz, Fach M697, Universit\"atsstra{\ss}e 10, 78457 Konstanz,
Germany.}

\begin{abstract}
In this work we comment some conclusions derived from the analysis
of recent proposals in the field of chaos-based cryptography. These
observations remark a number of major problems detected in some of
those schemes under examination. Therefore, this paper is a list of
what to avoid and to pay special attention to when considering chaos
as source of new strategies to conceal and protect information.
\end{abstract}
\end{frontmatter}

\section{Introduction}

The core of digital chaos-based cryptography is the selection of a
\emph{good} chaotic map for a given encryption scheme. Actually, the
presence of chaos does not guarantee the security of an encryption
algorithm \citep{kocarev01b}. A good digital cryptosystem based on
chaos should not be just the concomitance of a chaotic map and an
encryption architecture, but the result of their \emph{synergical}
association. Indeed, the quality of a chaotic map for cryptography
must be evaluated not just with considerations on its dynamic
properties, but also with considerations on the needs of the
sustaining encryption architecture. In other words, from a general
point of view it is not possible to design chaotic cryptosystems
satisfying the \index{chaotic-system-free property}
\emph{chaotic-system-free property} \citep[p.
30]{li:Dissertation2003} and, as a result, the selection of a
certain encryption scheme demands the selection of a group of
chaotic maps satisfying a certain set of dynamical properties.
Finally, digital chaos-based cryptography is implemented on
computers and thus the problem derived from finite-precision
computation must be evaluated and conveniently handled during the
design stage. This work illustrates the problems with three elements
involved in the design of digital chaos-based cryptosystems, i.e.,
the selection of a chaotic map (Sec.~\ref{sec:chaoticMap}), the
selection of an encryption architecture
(Sec.~\ref{sec:architecture}) and the implementation of the
encryption system (Sec.~\ref{sec:implementation}).

\section{Problems with the selection of the chaotic system}
\label{sec:chaoticMap}

\begin{problem}
\textbf{Definition of the key leading to non-chaotic behavior.} In
some chaos-based cryptosystems the control parameters of the
underlying chaotic systems are determined by the secret key. If the
link between the secret key and the control parameters is not
established carefully, then it is possible that the underlying
chaotic system evolves in an non-chaotic way, which further erodes
the confusion and diffusion properties required by the resulting
cryptosystem.
\end{problem}

The chaotic systems used as base of cryptosystems are defined in a
parametric way such that their dynamics depends on one or several
control parameters. Moreover, those chaotic systems are dynamical
systems which show a chaotic behavior for certain values of the
associated control parameter(s). Therefore, the design of a
cryptosystem based on any of those dynamical systems must be done by
guaranteeing the use of the set of values for the control
parameter(s) leading to chaos. Otherwise, the underlying dynamical
system associated to the cryptosystem (or encryption system) evolves
non-chaotically, which implies the reduction of the level of entropy
in the ciphertext (i.e., the output of the cryptosystem) and of the
influence on the ciphertext of a change in the plaintext (i.e., the
input of the cryptosystem). This problem is specially relevant when
the design of the cryptosystem is based on a dynamical system with
chaotic behavior only for a set of disjoint intervals of values of
the control parameter(s). This is the case of the logistic map and
the H\'enon map, which have been used in \citep{pareek03} and in
\citep{chee06} respectively without a thoroughly analysis of their
dynamics \citep{alvarez03b,arroyo07a}. As a conclusion, it is highly
advisable to use dynamical systems with chaotic behavior for all the
values of the control parameter(s). That is, \emph{robust chaotic
systems} \citep{RobustChaos98} should be used instead of nonrobust
ones.

\begin{problem}
\textbf{Nonuniform probability distribution fun\-ction.} In some
chaos-based encryption architectures the confusion and/or diffusion
properties depends on the probability distribution function of the
orbits derived from the selected chaotic systems. If that
distribution is not uniform and independent of the value(s) of
control parameter(s), then the quality of the diffusion process is
reduced.
\end{problem}

The iteration of a chaotic map can be used to generate pseudo-random
sequences to encrypt the plaintext. The encryption procedure could
be performed by different ways, but all of them demand the
equiprobability of all the states contained in the pseudo-random
sequences. If this requirement is not satisfied, then the
conditional entropy of the ciphertext with respect to the plaintext
may be large enough to leak information about relationships between
the output and the input of the target cryptosystem (see the entropy
attack in \citep{alvarez03a}). This effect is specially significant
for image encryption, as pointed out recently by
\cite{chengqingLi07} (see Fig.~5 of their paper). As a remedy,
chaotic maps with a uniform probability distribution function should
be selected as base of this kind of cryptosystems, being the family
of piecewise linear chaotic maps \citep{li05} a good option.

\begin{problem}
\textbf{Return map reconstruction.} The ciphertext of some
cryptosystems make it possible to reconstruct a return map of the
underlying chaotic system. If such a return map is meaningful, then
an attacker may be able to infer the value(s) of the control
parameter(s) that govern the evolution of the chaotic system.
\end{problem}

The most direct way to estimate the control parameter(s) from a
chaotic orbit is to plot $x_{n+1}$ versus $x_{n}$, which is actually
the chaotic map itself. If this representation shows a simple
function between $x_{n+1}$ and $x_{n}$, then it could be possible to
infer the control parameter. In \citep{skrobek08} a
chosen-ciphertext attack is used to build a discretized version of
the logistic map which further leads to the estimation of the
control parameter. One solution against this kind of attack is to
shuffle/truncate the chaotic orbit before using it for encryption,
which randomizes the plot of the the return map.

\section{Problems with the encryption architecture}
\label{sec:architecture}

\begin{problem}
\textbf{Bad definition of the ciphertexts.} A bad definition of the
ciphertext derived from a chaos-based cryptosystem could allow the
estimation of the initial condition(s) and/or the control
parameter(s) of the underlying chaotic system. This problem is
present in some chaos-based cryptosystems whose ciphertext is given
by fragments of orbits, sampled versions of the orbits, or
discretized versions of the orbits of the underlying chaotic
systems.
\end{problem}

\newcommand\mymatrix[1]{\bm{\mathrm{#1}}}

A $N$-dimensional discrete-time chaotic map is defined by the rule of
evolution
\begin{equation}
    \mymatrix{x}_{n+1}=f_{\mymatrix{\lambda}}(\mymatrix{x}_n),
\end{equation}
and, as a result, the ciphertext can not be the orbits of the map since it
may allow the estimation of $\mymatrix{\lambda}$ from $N+1$ or a bit
more consecutive units of ciphertext (see \citep{arroyo07c}). If the
invariant set of the chaotic map has a size dependent on the control
parameter(s), even sampled versions of the orbits may allow the
estimation of the control parameter(s). This is the case of the
cryptosystems reported in \citep{garcia02, pisarchik06} and cryptanalyzed
in \citep{alvarez03d, arroyo08a}. Finally, the theory of symbolic dynamics
can be used when the ciphertext allows to get the symbolic sequences of
the orbits of a chaotic map (see \citep{alvarez03b, arroyo08c}).

\begin{problem}
\textbf{Efficiency of the cryptosystem depending on the value of the
key.} If the encryption and decryption times depend on the key or a
sub-key, then a timing-attack can be performed to estimate the
(sub-)key.
\end{problem}

Some encryption architectures perform the transformation of the
plaintext into the ciphertext through several encryption rounds.
Additionally, in each encryption round a chaotic map is iterated $n$
times. Since the encryption and decryption time has to be constant
and independent of the value of the key, it is not a good practice
to select the number of encryption rounds and $n$ as part of the
key. Otherwise, a timing-attack based on the analysis of the
encryption and decryption time can be used for the partial
estimation of the secret key (see \citep{arroyo08a}), which is a
serious security flaw. Instead, the number of encryption rounds and
the number of iterations of the map should be public parameters of
the cryptosystem.

\begin{problem}
\textbf{Faulty derivation of the parameters of the chaotic system from the
key.} In some chaos-based cryptosystems the key is used to derive the
values of the parameters necessary to iterate a chaotic system and finally
encrypt the information. If this mapping implies a reduction of the key
space, i.e., that it is only used a subset of the possible values of those
parameters, then \modified{a} brute-force attack on the values of the
parameter could be much less demanding than the one on the secret key.
\end{problem}

One important step in the design of a chaos-based cryptosystem is to
decide what the key is. One possibility is to use the control
parameter(s) and the initial condition(s) of the underlying chaotic
system(s) as the secret key or as part of the secret key. Another
option is to establish the values of the control parameter(s) and
the initial condition(s) of the map(s) from the secret key through a
certain function. In this sense, it must be assured that the image
set of that function is the whole set of possible values of the
control parameter(s) and the initial condition(s). Otherwise, a
brute-force attack can be performed on the reduced space of control
parameter(s) and initial condition values with a lower computational
cost than the one on the key space. A cryptosystem with this problem
was introduced in \citep{pareek03} and was later cryptanalyzed in
\citep{alvarez03b}.

\begin{problem}
\textbf{Encryption procedure equivalent to a mapping only dependent
on the key.} If the transformation of the plaintext into the
ciphertext is determined by a procedure equivalent to a mapping only
dependent on the key, then known/chosen-plaintext attacks may be
performed to reconstruct the transformation procedure.
\end{problem}

In some encryption schemes the transformation of the plaintext into
the ciphertext is leading either by a procedure derived using only
the key, or by a sampling process on a sequence of values generated
using only the key. In those situations, it could be possible to
estimate either the key or to make up some function somehow
equivalent to the encryption procedure. For example, if the
encryption procedure consists of searching plaintexts in
pseudo-random sequences generated by iterating a chaotic map, since
the pseudo-random sequence remains unchanged unless the key is
modified, then it is possible to reconstruct the pseudo-random
sequence through a chosen-plaintext attack (see \citep{alvarez04a,
alvarez04b}). This problem also exists in those schemes where the
encryption procedure consists of a permutation-only stage which is
fixed unless the control parameter(s) and initial condition(s)
change, i.e., unless the the secret key is updated (see
\citep{Li:SPIC2008} for a general qualitative analysis of this
attack). As a conclusion, the encryption function that transforms a
unit of plaintext into a unit of ciphertext should depend on the key
and on the whole plaintext.
\section{Implementation problems}
\label{sec:implementation}

\begin{problem}
\textbf{Non-invertible encryption procedure.} The iteration of the chaotic
systems sustaining chaos-based cryptosystems implies working with real
numbers. Since the implementation of chaos-based cryptosystems is
done with finite precision arithmetic, round-off operations could lead to a
non-invertible encryption procedure.
\end{problem}

One critical point when working with dynamical systems and the analysis
of their dynamics is the selection of a right simulation framework. Indeed,
the computer-based analysis of dynamical systems could lead to some
conclusions different from those expected from theory. This divergence
also influences and conditions chaos-based cryptosystems. Thus, if the
characteristics and problems of finite-precision are not handled properly,
then it is possible that the orbits generated as base of encryption
procedure can not be regenerated exactly during the decryption stage
and, consequently, the original plaintext can not be recovered even when
the key is known. This problem is not only relevant for fixed-point
arithmetic but also for floating-point one. Indeed, the round-off quantization
errors could lead to the occurrence of a non-invertible function for
encryption and, as a result, the decryption process will be impossible (see
the cryptanalysis work in \citep{alvarez07, arroyo07a, arroyo08a,
solak08}).

\begin{problem}
\textbf{Dynamical degradation.} The implementation of chaotic
systems in finite precision in digital computers leads often to
dynamical properties completely different from the theoretical and
expected ones. If this deviation is not considered during the design
of chaos-based cryptosystems, it could imply a reduction of the
performance and even a compromise of the security of the resulting
cryptosystem.
\end{problem}

This problem is closely related to the previous one, although the
point of interest moves to degradation of dynamical properties of
the implemented chaotic system with respect to the theoretical
model. Consequently, the design of an encryption scheme using a
chaotic system must be done by considering its practical
implementation (not only the theoretical model). In
\citep{alvarez06b} some consequences of the dynamical degradation of
a chaotic map are shown in the context of cryptography, whereas in
\citep{li05} one can find a thorough analysis of the dynamical
degradation of a specific chaotic map and some ways to overcome this
problem.

\begin{problem}
\textbf{Lack of details in the description.} According to Kerckhoffs'
principle, the security of a cryptosystem can not be based on the secrecy
of its encryption and decryption procedures. Furthermore, the key of any
cryptosystem has to be easy to establish and to exchange, and the key
space must be defined in an explicit and clear way.
\end{problem}

The consecution of security through obscurity is something to avoid
when designing an encryption scheme. All the operations involved in
the encryption/decryption procedures must be verbosely explained,
and the secret key must be clearly specified along with an exact
estimation of the size of the key space. The security of the
cryptosystem must be only related to the difficulty of guessing the
key, and it can not depend on the lack of knowledge about the inner
operating of the encryption and decryption procedures. Moreover,
this lack of details implies a lack of security because without a
careful investigation of the whole cryptography community many
security holes might not be able to distinguished by the designers
themselves. Refer to \citep{arroyo07a} and \citep{li08} for a pair
of examples.

\section{Conclusions}

As a result of all the cryptanalysis work in the field of
chaos-based cryptography, we must conclude that the design of new
strategies of encryption using chaos must be based on a good
background on the theory of dynamical systems. In addition,
cryptanalytic knowledge about previous proposals and the
restrictions related to practical implementations on
finite-precision machines must be carefully studied and handled. A
cryptosystem is a chain composed of many links, whose security is
determined by the weakest link, and cryptanalysis is the art of
finding out the weakest link.

\bibliographystyle{ifacconf-harvard}

\begin{thebibliography}{24}
\expandafter\ifx\csname
natexlab\endcsname\relax\def\natexlab#1{#1}\fi
\expandafter\ifx\csname url\endcsname\relax
  \def\url#1{\texttt{#1}}\fi
\expandafter\ifx\csname urlprefix\endcsname\relax\def\urlprefix{URL
}\fi

\bibitem[{Alvarez and Li(2006)}]{alvarez06b}
Alvarez, G., Li, S., 2006. Breaking an encryption scheme based on
chaotic baker
  map. Physics Letters A 352~(1-2), 78--82.

\bibitem[{Alvarez et~al.(2007)Alvarez, Li, and Hernandez}]{alvarez07}
Alvarez, G., Li, S., Hernandez, L., 2007. Analysis of security
problems in a
  medical image encryption system. Computers in Biology and Medicine 37~(3),
  424--427.

\bibitem[{Alvarez et~al.(2003{\natexlab{a}})Alvarez, Montoya, and
  Pastor}]{alvarez03b}
Alvarez, G., Montoya, F., Pastor, G., 2003{\natexlab{a}}.
Cryptanalysis of a
  discrete chaotic cryptosystem using external key. Physics Letters A 319,
  334--339.

\bibitem[{Alvarez et~al.(2003{\natexlab{b}})Alvarez, Montoya, Romera, and
  Pastor}]{alvarez03d}
Alvarez, G., Montoya, F., Romera, M., Pastor, G.,
2003{\natexlab{b}}.
  Cryptanalysis of a chaotic secure communication system. Physics Letters A
  306~(4), 200--205.

\bibitem[{Alvarez et~al.(2003{\natexlab{c}})Alvarez, Montoya, Romera, and
  Pastor}]{alvarez03a}
Alvarez, G., Montoya, F., Romera, M., Pastor, G.,
2003{\natexlab{c}}.
  Cryptanalysis of an ergodic chaotic cipher. Physics Letters A 311, 172--179.

\bibitem[{Alvarez et~al.(2004{\natexlab{a}})Alvarez, Montoya, Romera, and
  Pastor}]{alvarez04b}
Alvarez, G., Montoya, F., Romera, M., Pastor, G.,
2004{\natexlab{a}}.
  Cryptanalysis of dynamic look-up table based chaotic cryptosystems. Physics
  Letters A 326, 211--218.

\bibitem[{Alvarez et~al.(2004{\natexlab{b}})Alvarez, Montoya, Romera, and
  Pastor}]{alvarez04a}
Alvarez, G., Montoya, F., Romera, M., Pastor, G.,
2004{\natexlab{b}}. Keystream
  cryptanalysis of a chaotic cryptographic method. Computer Physics
  Communications 156, 205--207.

\bibitem[{Arroyo et~al.(2008{\natexlab{a}})Arroyo, Alvarez, Li, Li, and
  Fernandez}]{arroyo08c}
Arroyo, D., Alvarez, G., Li, S., Li, C., Fernandez, V.,
2008{\natexlab{a}}.
  Cryptanalysis of a new chaotic cryptosystem based on ergodicity.
  http://arxiv.org/abs/0806.3183.

\bibitem[{Arroyo et~al.(2008{\natexlab{b}})Arroyo, Alvarez, Li, Li, and
  Nunez}]{arroyo07a}
Arroyo, D., Alvarez, G., Li, S., Li, C., Nunez, J.,
2008{\natexlab{b}}.
  Cryptanalysis of a discrete-time synchronous chaotic encryption system.
  Physics Letter A 372~(7), 1034--1039.

\bibitem[{Arroyo et~al.(2008{\natexlab{c}})Arroyo, Li, Li, and
  Alvarez}]{arroyo07c}
Arroyo, D., Li, C., Li, S., Alvarez, G., 2008{\natexlab{c}}.
Cryptanalysis of a
  computer cryptography scheme based on a filter bank. Chaos, Solitons and
  Fractals, In Press.

\bibitem[{Arroyo et~al.(2008{\natexlab{d}})Arroyo, Rhouma, Alvarez, Li, and
  Fernandez}]{arroyo08a}
Arroyo, D., Rhouma, R., Alvarez, G., Li, S., Fernandez, V.,
2008{\natexlab{d}}.
  On the security of a new image encryption scheme based on chaotic map
  lattices. Chaos 18, Art. No. 033112, 7 pages.

\bibitem[{Banerjee et~al.(1998)Banerjee, Yorke, and Grebogi}]{RobustChaos98}
Banerjee, S., Yorke, J.~A., Grebogi, C., 1998. Robust chaos.
Physical Review
  Letters 80, 14.

\bibitem[{Chee and Xu(2006)}]{chee06}
Chee, C.~Y., Xu, D., 2006. Chaotic encryption using dicrete-time
synchronous
  chaos. Physics Letters A 348~(3-6), 284--292.

\bibitem[{Garc\'ia and Jim\'enez(2002)}]{garcia02}
Garc\'ia, P., Jim\'enez, J., 2002. Communication through chaotic map
systems.
  Physics Letters A 298~(1), 35--40.

\bibitem[{Kocarev(2001)}]{kocarev01b}
Kocarev, L., 2001. Chaos-based cryptography: A brief overview. IEEE
Circuits
  and Systems Magazine 1~(2), 6--21.

\bibitem[{Li et~al.(2007)Li, Li, Alvarez, Chen, and Lo}]{chengqingLi07}
Li, C., Li, S., Alvarez, G., Chen, G., Lo, K.-T., 2007.
Cryptanalysis of two
  chaotic encryption schemes based on circular bit shift and {XOR} operations.
  Physics Letters A 369, 23--30.

\bibitem[{Li et~al.(2008{\natexlab{a}})Li, Li, Chen, and Halang}]{li08}
Li, C., Li, S., Chen, G., Halang, W.~A., 2008{\natexlab{a}}.
Cryptanalysis of
  an image encryption scheme based on a compound chaotic sequence. Image and
  Vision Computing. Article in Press.

\bibitem[{Li(2003)}]{li:Dissertation2003}
Li, S., June 2003. Analyses and new designs of digital chaotic
ciphers. Ph.D.
  thesis, School of Electronic and Information Engineering, Xi'an Jiaotong
  University, Xi'an, China, available online at
  \url{http://www.hooklee.com/pub.html}.

\bibitem[{Li et~al.(2005)Li, Chen, and Mou}]{li05}
Li, S., Chen, G., Mou, X., 2005. On the dynamical degradation of
digital
  piecewise linear chaotic maps. International Journal on Bifurcation and Chaos
  15~(10), 3119--3151.

\bibitem[{Li et~al.(2008{\natexlab{b}})Li, Li, Chen, Bourbakis, and
  Lo}]{Li:SPIC2008}
Li, S., Li, C., Chen, G., Bourbakis, N.~G., Lo, K.-T.,
2008{\natexlab{b}}. A
  general quantitative cryptanalysis of permutation-only multimedia ciphers
  against plaintext attacks. Signal Processing: Image Communication 23~(3),
  212--223.

\bibitem[{Pareek et~al.(2003)Pareek, Patidar, and Sud}]{pareek03}
Pareek, N.~K., Patidar, V., Sud, K.~K., 2003. Discrete chaotic
cryptography
  using external key. Physics Letters A 309, 75--82.

\bibitem[{Pisarchik et~al.(2006)Pisarchik, Flores-Carmona, and
  Carpio-Valadez}]{pisarchik06}
Pisarchik, A.~N., Flores-Carmona, N.~J., Carpio-Valadez, M., 2006.
Encryption
  and decryption of images with chaotic map lattices. Chaos 16~(3), Art. No.
  033118.

\bibitem[{Skrobek(2008)}]{skrobek08}
Skrobek, A., 2008. Approximation of a chaotic orbit as a
cryptanalytical method
  on {Baptista}'s cipher. Physics Letters A 372~(6), 849--859.

\bibitem[{Solak and \c{C}okal(2008)}]{solak08}
Solak, E., \c{C}okal, C., 2008. Comment on ``encryption and
decryption of
  images with chaotic map lattices''. Chaos 18~(3), 038101.

\end{thebibliography}

\end{document}